\documentclass[aps,prb,preprint]{revtex4-1}
\usepackage{graphicx,color}

\usepackage{dcolumn}
\usepackage{bm}
\usepackage{amsmath}
\usepackage{ulem}
\usepackage{wasysym}%

\begin{document}

%\title{Tuning the coupling of Bi$_2$Se$_3$-based topological insulators  to its ferromagnetic phase: MBE-grown ($n$-QLs Bi$_2$Se$_3$/Mn-Bi$_2$Se$_3$) heterostructures}

\title{Topological insulator homojunctions including magnetic layers: the example of n-p type ($n$-QLs Bi$_2$Se$_3$/Mn-Bi$_2$Se$_3$) heterostructures}

\author{M. Vali\v{s}ka}  
\affiliation{Department of Condensed Matter Physics, Charles University, 120 00 Praha 2, Czech Republic}

%\author{R. Tarasenko}  
%\affiliation{Department of Condensed Matter Physics, Charles University, CZ-120 00 Praha 2, Czech Republic}

\author{J. Warmuth}  
\affiliation{Department of Physics, University of Hamburg, D-20355 Hamburg, Germany}

\author{M. Michiardi}  
\affiliation{Department of Physics and Astronomy,
Interdisciplinary Nanoscience Center (iNANO),
University of Aarhus, 8000 Aarhus C, Denmark}

\author{M. Vondr\'a\v{c}ek}  
\affiliation{Institute of Physics, ASCR, 182 21 Praha 8, Czech Republic}

\author{A. S. Ngankeu}  
\affiliation{Department of Physics and Astronomy,
Interdisciplinary Nanoscience Center (iNANO),
University of Aarhus, 8000 Aarhus C, Denmark}

\author{V. Hol\'y}  
\affiliation{Department of Condensed Matter Physics, Charles University, 120 00 Praha 2, Czech Republic}

%\author{K. Carva}  
%\affiliation{Department of Condensed Matter Physics, Charles University, CZ-120 00 Praha 2, Czech Republic}

\author{V. Sechovsk\'y}  
\affiliation{Department of Condensed Matter Physics, Charles University, 120 00 Praha 2, Czech Republic}

\author{G. Springholz}  
\affiliation{Institute of Semiconductor and Solid State Physics, Johannes Kepler University, 4040 Linz, Austria}

\author{M. Bianchi}  
\affiliation{Department of Physics and Astronomy,
Interdisciplinary Nanoscience Center (iNANO),
University of Aarhus, 8000 Aarhus C, Denmark}

\author{J. Wiebe}  
\affiliation{Department of Physics, University of Hamburg, D-20355 Hamburg, Germany}

%\author{R. Wiesendanger}  
%\affiliation{INF, University of Hamburg, Germany}

\author{P. Hofmann}  
\affiliation{Department of Physics and Astronomy,
Interdisciplinary Nanoscience Center (iNANO),
University of Aarhus, 8000 Aarhus C, Denmark}

\author{J. Honolka}
\affiliation{Institute of Physics, ASCR, 182 21 Praha 8, Czech Republic}

\date{\today}

\begin{abstract}
Homojunctions between Bi$_2$Se$_3$ and its Mn-doped phase are investigated 
 as a sample geometry to study the influence of spin degrees of freedom on topological insulator properties.
 $n$ quintuple layers (QLs) of Bi$_2$Se$_3$ are grown ontop of Mn-doped Bi$_2$Se$_3$ by molecular beam epitaxy for $0\le n \le 30\,$QLs, allowing to unhamperedly monitor the development of electronic and topological properties by surface sensitive techniques like angle resolved photoemission spectroscopy. With increasing $n$, a Mn-induced gap at the Dirac point is gradually filled in an "hourglass" fashion to reestablish a topological surface state at $n \sim 9\,$QLs. Our results suggest a competition of upwards and downwards band bending effects due to the presence of an n-p type interface, which can be used to tailor topological and quantum well states independently.   
%The domain formation on a micrometer scale typical for MBE growth under non-equilibrium growth conditions seems to leave the topological state intact, suggesting that the mean free path of topological states is considerably smaller than a micrometer. 

\end{abstract}


\maketitle

\begin{figure}
\center \includegraphics[width=85mm]{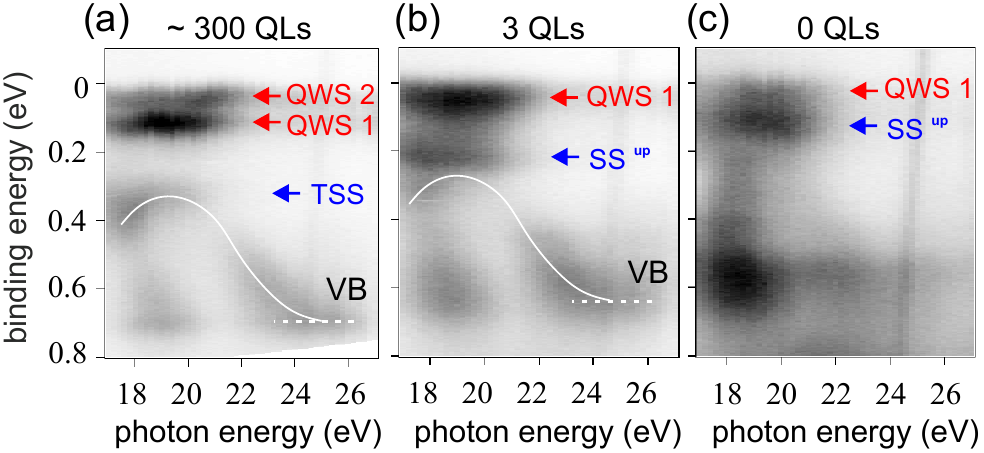}
\caption{Photon energy dispersion at the $\bar{\Gamma}$ point for $n\sim 300$ (a), $n=3$ (b), and $n=0$ (c) (series $\mathcal{B}$).
Starting as a TSS, the SS splits into SS$^{\text{up/down}}$ and SS$^{\text{up}}$ shifts up in energy keeping its non-dispersive 2D nature. Additional QWSs 1 and 2 are indicated by arrows. Dispersing valence bands (VB) are highlighted by contiuous lines as a guide to the eye. 
}
\label{ARPES-dispersion}
\end{figure}

\begin{figure}
\center \includegraphics[width=87mm]{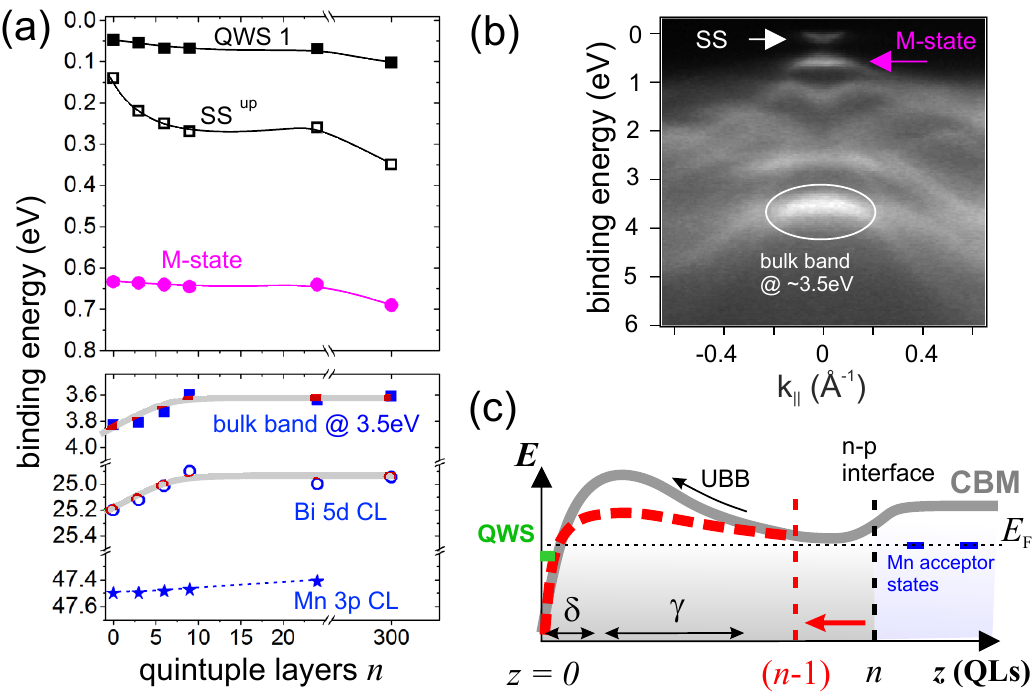}
\caption{(a) Energy shifts of QWS, SS, and M-state (top), as well as CLs and a bulk band at $3.5\,$eV (bottom) versus overgrown QLs $n$. Delocalized band states are evaluated at $\bar{\Gamma}$. (b) Typical ARPES image showing a bulk band with maximum at $3.5\,$eV. (c) Qualitative model of the influence of the n-p interface on band bending effects. 
}
\label{ARPES-trend}
\end{figure}

The material class of topological insulators (TIs) is presently in the focus of research, due to its fascinating fundamental physics but also its potential for future spintronic applications~\cite{HasanOverview, Qi2011}. Experimentally, the TI phase was first observed in 2D HgTe/HgCdTe quantum well systems, possessing spin-polarised helical 1D edge states~\cite{Koenig2007, Bernevig2006}. Later, chalcogenide-type materials like Bi$_2$Te$_3$ and Bi$_2$Se$_3$ with large spin-orbit coupling were shown to host 2D topological surface states (TSSs), defined by a Dirac-like dispersion, which locks momentum solutions $+k$ and $-k$ to opposite spin~\cite{Pesin2012}. 
The intimate link between time reversal symmetry and TI properties soon raised the fundamental question how magnetic degrees of freedom influence TI properties. Stable local moments with $M_{\perp} \neq 0$ perpendicular to the surface should break time reversal symmetry and are predicted to cause an energy gap at the Dirac point (DP)~\cite{Liu2009, Henk2012}.\\ 
Strategies to couple magnetic moments to the TSS are so far devided into surface concepts (magnetic adatoms) and bulk strategies (magnetic dopants). Surface concepts included adsorbed magnetic single atoms~\cite{Honolka2012, Scholz2012, Eelbo2014, Sessi2014} and superparamagnetic clusters~\cite{Wray2011}. The disadvantage of surface approaches are the lack of stable moments in the impurity limit (fluctuating moments with $\langle M\rangle=0$) and the obstructed access to the TSS at larger coverages by experimental techniques such as surface sensitive angle-resolved photoelectron spectroscopy (ARPES). Bulk strategies promise larger potential since a stable ferromagnetic phase can be induced within the TI bulk material itself, potentially influencing the TSSs within their decay length into the bulk over several nanometers~\cite{Can-Li2010}. For chalcogenide based materials %such as magnetically doped Bi$_2$Te$_3$ or Bi$_2$Se$_3$ 
theory predicts several possible magnetic interaction mechanisms~\citep{Biswas2010, Abanin2011, Efimkin2014, Choi2012} and ferromagnetic ordering was reported e.g. for 3$d$ elements Fe, Mn, and Cr in bulk Bi$_2$Te$_3$ or Bi$_2$Se$_3$ samples~\cite{Su-Yang2012, Zhang2012b, Bardeleben2013} with Curie temperatures $T_c$ ranging from $1\,$K to $40\,$K.
Although bulk concepts are vividly studied, reports on the influence of ferromagnetic bulk phases on TSSs are contradictive and  controversively dicussed~\cite{Su-Yang2012, Ruzicka2015}. Moreover, assignments to existing theory on magnetic ordering is difficult, since 3$d$ atoms tend to integrate into the host matrix in various ways, strongly depending on sample preparation parameters.\\
%from substitutional 3d$_\text{Bi}$ to interstitial sites within a quintuple layer (QL) and in the van-der-Waals (vdW) gap between QLs. \\
\indent In this work, we follow a different approach based on layered Bi$_2$Se$_3$ heterostructures with a buried magnetic phase. Using molecular beam epitaxy (MBE), an increasing number $n$ of pure Bi$_2$Se$_3$ quintuple layers (QLs) is grown ontop of Mn-doped Bi$_2$Se$_3$. ARPES and X-ray photoelectron spectroscopy (XPS) were done in Aarhus at the synchrotron ASTRID 2~\cite{Hoffmann2004aa} and in Prague using an Omicron NanoESCA instrument with laboratory light sources. Our band structure data suggest a gradual decoupling of the Bi$_2$Se$_3$ TSS from influences of the buried magnetic phase, and non-monotoneous band bending (BB) profiles caused by the proximity of the n-p type interface with Mn-doped Bi$_2$Se$_3$. Our results stress the potential of such systems to study time reversal symmetry breaking influences on TSSs under well-defined pertubative conditions exploiting band engineering strategies.\\ 
%%%%%%%%%%%%%%%%%%%%
%%%%%%%%%%%%%%%%%%%%
\begin{figure}
\center \includegraphics[width=80mm]{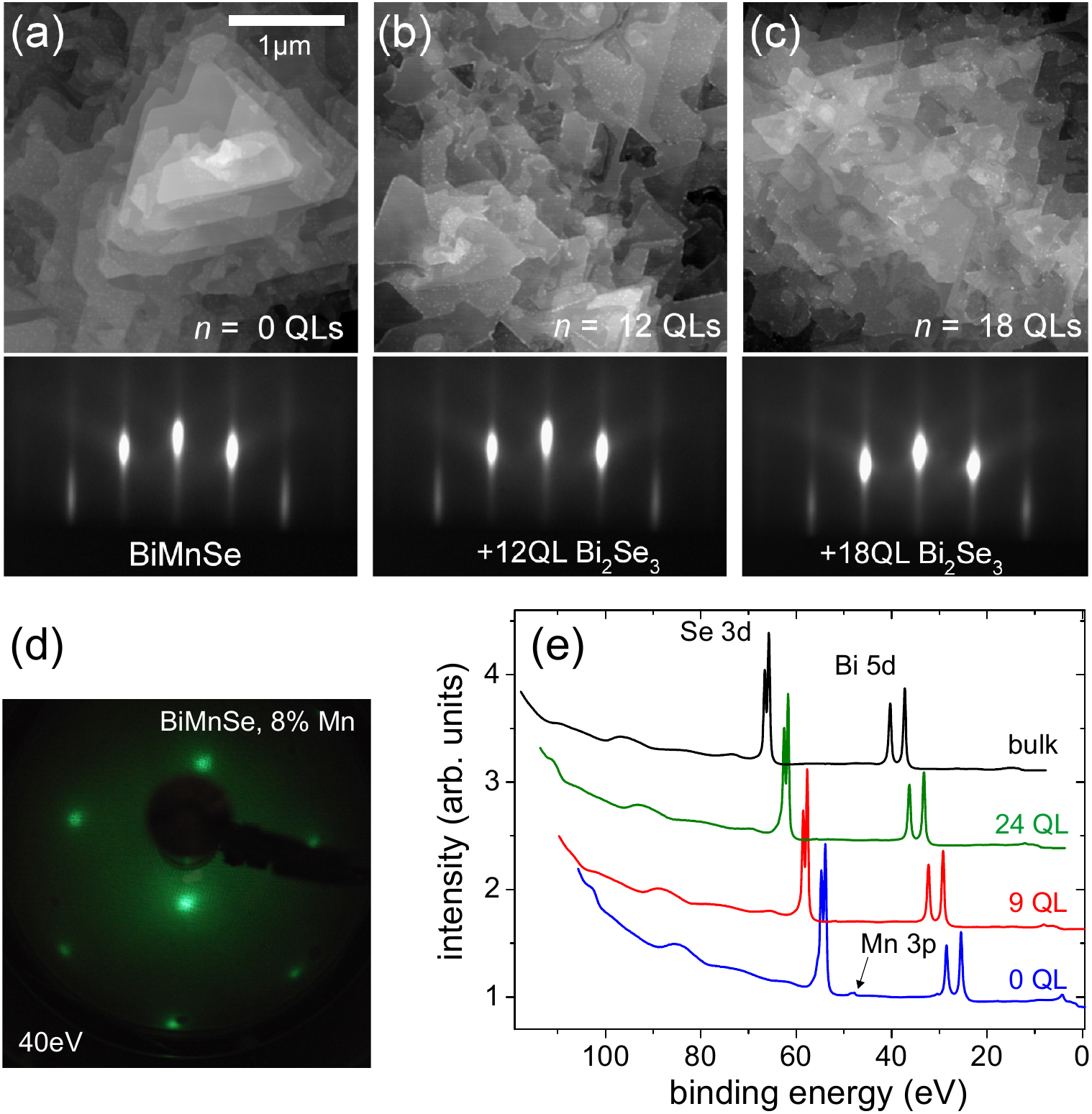}
\caption{
(a)-(c): Terraced surfaces of ($n$-QLs Bi$_2$Se$_3$/Mn-Bi$_2$Se$_3$) heterostructures for different $n$. Top and bottom: AFM topography (dimensions $3\,\mu\text{m} \times 3\,\mu\text{m}$) and RHEED, respectively.
(d) Typical LEED image at $E=40\,$eV after decapping. (e) Bi and Se core level XPS for $n=0, 9, \text{ and}\, 24\,$QLs in comparison to melt-grown Bi$_2$Se$_3$. The photon energy was $h\nu=100\,$eV and spectra are artificially shifted in energy for better visibility.
}
\label{fign1}
\end{figure}
%%%%%%%%%%%%%%%%%%%%
%%%%%%%%%%%%%%%%%%%%
Bi$_2$Se$_3$-based systems were epitaxially grown on insulating BaF$_2$(111) 
%at temperatures between 360$^{\circ}$C and 380$^{\circ}$C. 
and the structural order of the material was monitored by reflection high energy electron diffraction (RHEED). Since the (0001) basal plane of the trigonal Bi$_2$Se$_3$ lattice is nearly lattice-matched to BaF$_2$(111), two-dimensional layer growth with good crystalline quality can be obtained. For details on the growth procedure see Ref.~\cite{Caha2013}. 
Mn-doped Bi$_2$Se$_3$ thin films of different thickness were grown at $T=380^{\circ}$C under Se-rich conditions to support 3d$_{\text Bi}$ substitutional implementation with respect to e.g. interstitial defects~\cite{Can-Li2010}. Additional quintuple layers of pure Bi$_2$Se$_3$ were grown continuously ontop the Mn-doped phase at slightly lower temperatures $T=330^{\circ}$C to form $n$-QLs Bi$_2$Se$_3$/Mn-Bi$_2$Se$_3$ heterostructures. In this work data of two sample series $\mathcal{A}$ ($n=0, 4, 8, \text{and } 12$) and $\mathcal{B}$ ($n=0, 3, 6, 9, 24, \text{and } 30$) on respective Mn-doped Bi$_2$Se$_3$ films of 500nm and 300nm thickness are shown.\\
The epitaxial heterostructures were finally protected against oxidation by amorphous Se cap layers, which allows to recover large, atomically flat terraces by an in-situ decapping procedure under ultra-high vacuum conditions (see detailed information supplementary, Fig.~S1). Se desorption was monitored by mass spectrometry, and successful decapping was verified in Se and Bi XPS core level (CL) data for each sample shown in this work. \\
%%%%%%%%%%%%%%%%%%%%
%%%%%%%%%%%%%%%%%%%%
\begin{figure}
\center \includegraphics[width=86mm]{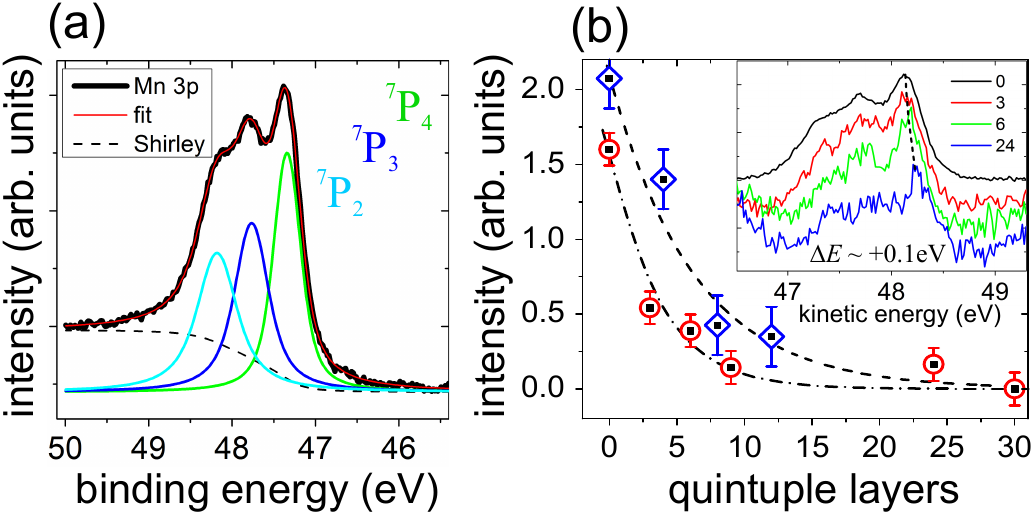}
\caption{
(a) Mn 3$p$ XPS data for $n=0$ ($h\nu=130\,$eV). The fit corresponds to a $^{7}P_4$, $^{7}P_3$, and $^{7}P_2$ multiplet with Shirley background. (b) Mn 3$p$ ($^{7}P_4$) and 2$p_{3/2}$ peak intensities versus $n$ for sample series $\mathcal{A}$ ($\diamond$, $h\nu=1.5\,$keV) and $\mathcal{B}$ ($\circ$, $h\nu=100\,$eV), respectively. The Lambert-Beer fits correspond to photoelectron escape depths of $3.5\,$QLs $\sim 3.3\,$nm and $6.2\,$QLs $\sim 6\,$nm. 
Inset: rescaled Mn 3$p$ spectrum versus kinetic energy $E_{\text{kin}}$, shifting by $+0.1\,$eV with $n$. 
}
\label{Mn-XPS}
\end{figure}
%%%%%%%%%%%%%%%%%%%%
%%%%%%%%%%%%%%%%%%%%
From bulk sensitive energy dispersive X-ray spectroscopy (EDX) maps with a lateral resolution of about 1$\,\mu$m we can exclude significant variations of the Mn concentration or crystal precipitations on this length scale and beyond. Estimations of Bi, Se, and Mn concentrations are summarized in Table~\ref{EDX-table} for two sample series $\mathcal{A}$ and $\mathcal{B}$, respectively. Comparing the stoichiometry of films with and without Mn, we find that the Bi/Se ratio 2/3 is not significantly altered by Mn implementation. At Mn concentrations $x_{\text Mn} \sim 10\%$ we find Curie temperatures $T_{\text{C}}$ of about 5K in SQUID measurements. Bulk sensitive X-ray diffraction studies, 
%as well as surface sensitive electron backscattering diffraction (EBSD) with lateral resolution of about $20\,$nm 
moreover, prove a high degree of order in the film. For details we refer the reader to our separate work focused on bulk properties~\cite{Tarasenko2016}.\\
%%%%%%%%%%%%%%%%%%%%
%%%%%%%%%%%%%%%%%%%%
\begin{table}[h!]
\caption{Mn-Bi$_2$Se$_3$ film properties (without overlayers) derived from EDX and SQUID. The stoichiometry is given in units of atomic percent with an error of $\pm 0.5\,$\%.}
      \begin{tabular}{cccccc}
        \hline
          & $x_{\text{Bi}}\,[\%]$  & $x_{\text{Se}}$ [\%]  & $x_{\text{Bi}}/x_{\text{Se}}$ & $x_{\text{Mn}}$[\%] & $T_{\text{C}}$ [K]\\ \hline
        Series $\mathcal{A}$ ($500\,$nm) & 36.6 & 51.3 & 0.71 &12.1 & 5.4\\
        Series $\mathcal{B}$ ($300\,$nm) & 34.6 & 51.1  & 0.68 & 13.8 & 5.2\\
        % 300nm Bi$_2$Se$_3$ & 36.6  & 50.9  & 0.72 & 0 & --\\
        Melt-grown Bi$_2$Se$_3$ & 40.5  & 59.5 & 0.68 & 0 & --\\\hline
      \end{tabular}
      \label{EDX-table}
\end{table}
%%%%%%%%%%%%
%%%%%%%%%%%%%%
Fig.~\ref{fign1} summarizes the overgrowth properties of ($n$-QLs Bi$_2$Se$_3$/Mn-Bi$_2$Se$_3$) heterostructures with different number of QLs. AFM images (a)-(c) at $n=0, 12,$ and $18\,$ show material-typical pyramidal growth during MBE.
% as reported before for homogeneous bulk samples (Ref. Springholz). 
We find that the height corrugation at $n=12$ and $n=1$ is similar to the homogeneous phase with $n=0$, and triangular terraces show $180^{\circ}$ mirrored geometries typical for twinning in chalcogenide TIs under the influence of 3$d$ metal dopants. In addition, the average terrace width is reduced compared to pure MBE-grown Bi$_2$Se$_3$.
The corresponding RHEED patterns do not significantly change with $n$, indicating that despite the increased corrugation, the crystal growth conditions do not change dramatically at the interface between Mn-doped and undoped Bi$_2$Se$_3$. LEED images in Fig.~\ref{fign1}(d) and XPS Se~3$d_{3/2, 5/2}$ and Bi~5$d_{3/2}$ core level spectra shown in (e) confirm well-defined (111)-oriented and chemically ordered surfaces after decapping.\\
%%%%%%%%%%%%%%%%%%%%
%%%%%%%%%%%%%%%%%%%%
\begin{figure*}
\center \includegraphics[width=178mm]{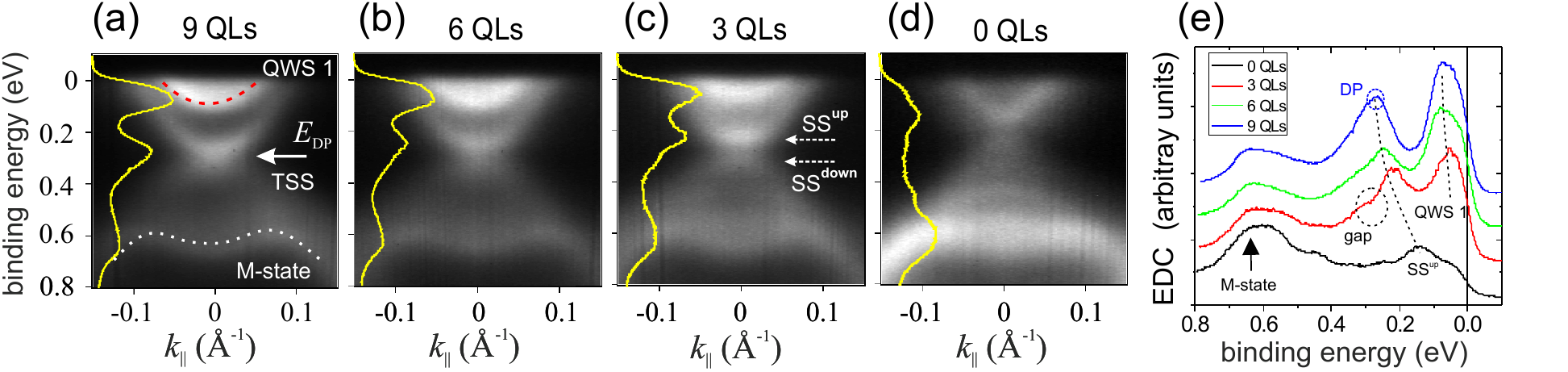}
\caption{(a)-(d) ARPES data of ($n$-QLs Bi$_2$Se$_3$/Mn-Bi$_2$Se$_3$) heterostructures along the $\bar{\text{M}}-\bar{\Gamma}-\bar{\text{M}}$ direction at photon energy $h\nu=18\,$eV and $T=100\text{K}$ (sample series $\mathcal{B}$). Overlaid are energy distribution curves (EDC) derived from the $k$-interval $[-0.01\,\text{\AA}^{-1}, 0.01\,\text{\AA}^{-1}]$ at respective binding energies. (e) shows all EDCs for direct comparison.}
\label{ARPES}
\end{figure*}
%%%%%%%%%%%%%%%%%%%%
%%%%%%%%%%%%%%%%%%%%
Highly-resolved XPS allows to detect chemical shifts in Mn, Bi and Se CLs, which depend on the atom`s chemical environments within the XPS sampling depth $\lambda$.
% Fig.~\ref{Mn-XPS} summarizes XPS data on Mn core levels measured for different $n$ were measured at photon energies $h\nu=100\,$eV with highest surface sensitivity.
First we are interested in the electronic ground state of Mn dopants itself. The Mn 3$p$ core level spectrum in Fig.~\ref{Mn-XPS}(a) for $n=0$ exhibits a sharp multiplet structure, suggesting a well defined Mn electronic configuration in the Bi$_2$Se$_3$ host. This is in contrast to a metallic Mn phases where a broad almost featureless 3$p$ spectrum is observed at same photon energies due to solid state broadening~\cite{Noh2015, vondemBorne2000}. The comparison proves that Mn atoms are not significantly clustered in our samples. Instead, the spectrum can be perfectly described by a ($^{7}P_4$, $^{7}P_3$, $^{7}P_2$) multiplet of a Mn $d^5$ ground state, which under the influence of core hole interactions splits in binding energy (BE) with the according intensities (see Fig.~\ref{Mn-XPS}(a)). The width of the three Voigt components increases with BE, as expected from increasing super-Coster-Kronig decay processes caused by the large overlap of the Mn 3$p$ with the 3$d$ wavefunctions~\cite{MartinsOverview2006}. For a direct comparison of the line shape evolution with $n$, the inset of Fig.~\ref{Mn-XPS}(b) presents the scaled Mn 3$p$ signals, which shows that the width of the Mn multiplet does not change significantly. As we will discuss in more detail below, the gradual and rigid shifts of the triplet structure by up to $\Delta E=+0.1\,$eV towards higher kinetic energies suggests upwards band bending (UBB) effects within the probed buried Mn-doped Bi$_2$Se$_3$ phase with inreasing $n$.\\
The XPS probing depth $\lambda$ can be directly estimated from the Mn XPS peak intensities of sample series $\mathcal{A}$ and $\mathcal{B}$ for photon energies $h\nu=1480\,$eV and $h\nu=100\,$eV, respectively, as summarized in Fig.~\ref{Mn-XPS}(b). Mn 2$p$ and 3$p$ intensities decay approximately exponentially with increasing $n$ and the fits using Lambert-Beer's law $I(n)=I_0\times e^{-n/\lambda}$ correspond to effective electron mean free paths $\lambda_{\mathcal{A}} = (6.2\pm 0.5)\,\text{QLs} \sim 5.9\,$nm and $\lambda_{\mathcal{B}} = (3.5\pm 0.5)\,\text{QLs} \sim 3.3\,$nm. $\lambda_{\mathcal{A}}>\lambda_{\mathcal{B}}$ is expected due to higher kinetic energy of electrons excited with $h\nu=1487\,$eV. The value $\lambda_{\mathcal{B}}$ is larger than expected for the given kinetic energy.\\
%%%%%%%%%%%%%%%%%%%%%%%%%%%%%%%%%%%%
The heterostructure geometry allows the unobstructed study of the TSS under the influence of the buried magnetic phase. Fig.~\ref{ARPES}(a)-(d) summarizes the ARPES spectra of the sample series $\mathcal{B}$ measured at $T=100\,$K and and photon energy $h\nu=18\,$eV. We stress that precise cuts through the $\bar{\Gamma}$ point were extracted from data sets covering the full Fermi surface range.\\
The ARPES cuts along $\bar{\text{M}}-\bar{\Gamma}-\bar{\text{M}}$ reveal clear trends with decreasing $n$. 
Starting from large numbers of QLs, at $n=9$ a typical n-doped Bi$_2$Se$_3$ TSS is observed in Fig.~\ref{ARPES}(a) with a DP located at $E_{\text{DP}} = 0.3\,$eV. %and $k_{\parallel}=0$. 
In addition, a parabolic band is visible close to $E_{\text{F}}$ and M-shaped states dominate at $\sim 0.6\,$eV. Both are well-known features for n-doped Bi$_2$Se$_3$, originating from conduction and valence band states, respectively~\cite{Bianchi2011}. Fig.~\ref{ARPES-dispersion} proves that in our case the parabolic states close to $E_{\text{F}}$ do not disperse with photon energy, typical for 2D quantum well states (QWS) formed in a near-surface downwards band bending (DBB) potential.\\
% formed under the influence of strong downwards band bending (DBB) at the surface~\cit{Bahramy2012}.
%%%%%%%%%%%%%%%%%%%%
%%%%%%%%%%%%%%%%%%%%
%%%%%%%%%%%%%%%%%%%%
%%%%%%%%%%%%%%%%%%%%
Reducing the number of QLs, the TSS gradually becomes massive forming a gap at the DP, limited at the top and bottom by the bands SS$^{\text{up}}$ and SS$^{\text{down}}$ as indicated in Fig.~\ref{ARPES}(c). We emphasize that ARPES was done in the paramagnetic phase $T>T_{\text{C}}$. Fig.~\ref{ARPES-dispersion}(b) shows that SS$^{\text{up}}$ does not disperse with photon energy and thus inherits the TSS`s 2D nature as expected. Finally, Fig.~\ref{ARPES}(d) shows that for $n=0$, disctinct resonances appear in a seemingly increased gap between the M-state and SS$^{\text{up}}$.\\
Trends of intensity around the $\bar{\Gamma}$ point are best visible in energy distribution curves (EDCs) averaged in the range [$-0.01\,\text{\AA}^{-1}$, $+0.01\,\text{\AA}^{-1}$], which are plotted in the respective ARPES images and for direct comparison in Fig.~\ref{ARPES}(e). Starting as an intact TSS at $n=9$, a gap at $E_{\text{DP}}$ manifests itself as a dip at in the EDC, best visible for $n=3$. 
%While the development of SS$^{\text{down}}$ is less visible in EDCs, 
The development of SS$^{\text{up}}$, QWS 1 and M-state with $n$ are shown in the top panel of Fig.~\ref{ARPES-trend}(a).
SS$^{\text{up}}$ shifts up towards lower BEs by more than $0.10\,$eV between $n=9$ and $n=0$, while QWS and M-state shift by less than $0.05\,$eV. 
% Comment: This is also visible in the high-energy ARPES with hv=60.5eV. Also here the shift of M-states is less that 0.05, although the resolution gives a binding energy value (0.57 /pm 0.03) at Gamma, smaller than the low photon energy value
The parallel development of QWS and M-states with $n$ suggests subtle changes of near-surface DBB, similar to those induced e.g. by adsorbates on bulk Bi$_2$Se$_3$~\cite{Bianchi2012}. In contrast, the much stronger shift of SS$^{\text{up}}$ in the range $n \le 6\,$QLs proves an independent development for states of topological origin, e.g. due to subsurface Mn impurities creating a gap and introducing resonances around the DP~\cite{Black2012} as observed in our ARPES data. We briefly mention that below $n = 6\,$QLs the overlap of degenerate front and backside TSS wavefunctions in free standing Bi$_2$Se$_3$ films should create a gap~\cite{Park2015}, however, this effect can be disabled due to the Mn-doped Bi$_2$Se$_3$ interface, which induces onesided energy shifts~\cite{Eremeev201530}.\\ 
The interpretation of our data is complicated by the fact that in contrast to SSs and QWSs, CLs in XPS move to lower BEs with $n$ as discussed above for Mn 3$p$, and observed much more distinctively for Bi 5$d$ and Se $3d$ CLs in both sample series $\mathcal{A}$ and $\mathcal{B}$ (see Fig.~\ref{ARPES-trend}(a), bottom panel and Supplementary Fig.~S2). For a better understanding we measured ARPES up to BEs of $6\,$eV as shown in Fig.~\ref{ARPES-trend}(b). A characteristic Bi$_2$Se$_3$ bulk band maximum at BE $\sim 3.5\,$eV is visible~\cite{Zhang2010}, which shifts strictly parallel to the localized Bi and Se CLs with increasing $n$ (see Fig.~\ref{ARPES-trend}(a), lower panel), suggesting UBB in a surface volume according to our probing depth $\lambda \sim\,$ 3-4~QLs.
UBB effects of $50-200\,$meV on length scales larger than 3-4~QLs have indeed been predicted for heavily and intermediately n-doped Bi$_2$Se$_3$~\cite{Veyrat2015, Foerster2015} in the absence of surface doping as a consequence of bulk charge transfer from donor states (e.g. Se-vacancy states) into available SSs. For our heterostrucures we seem to be able to tune this effect by the parameter $n$. From the fact that Mn CL shifts are considerably smaller, we infer that BB happens mostly in the Bi$_2$Se$_3$ overlayers. Taking the shifts of all states into account we propose that near-surface DBB only affect QWS in a very small subsurface range $z\le\delta$,
while UBB governs deeper layers $\delta \le z\le \gamma$ as shown in the conduction BB schemes in Fig.~\ref{ARPES-trend}(c) for the cases $n$ and $(n-1)\,$QLs. Such a band bending model is supported by recent theoretical predictions~\cite{Park2013}. This model, which assumes Mn to create midgap or shallow p-type levels~\cite{Zhang2014, Choi2012}
%, supported by our resistivity data $\rho(T)$, showing negative slopes $d\rho/dT < 0$ below $\sim$20K for Mn-Bi$_2$Se$_3$ films with $x_{\text{Mn}}>5\%$. 
and thus a p-n-interface with Bi$_2$Se$_3$ overlayers, qualitatively explains both the independent spectroscopic trends of QW, surface, and bulk electron states, as well as the different degrees of UBB effects observed in buried Mn and overlayer Bi/Se CLs.\\
\indent In summary, we studied MBE-grown layered Bi$_2$Se$_3$ heterostructures with a buried Mn-doped phase.
The geometry of $n$ quintuple layers (QLs) Bi$_2$Se$_3$ ontop of Mn-doped Bi$_2$Se$_3$ allows to unhamperedly monitor the development of surface electronic properties, which reveals a gradual recovery of a gapped topological surface state with increasing $n$, but also strong band bending effects due to the surface proximity of the heterostructure interface. Our results stress the potential of such sample geometries as a model system to study magnetic influences on topological insulator phases under well-defined pertubative conditions with the option to exploit band engineering techniques. 
%%%%%%%%%%%%%%%%%%%%
%%%%%%%%%%%%%%%%%%%%
%%%%%%%%%%%%%%%%%%%%
%%%%%%%%%%%%%%%%%%%%

\begin{acknowledgments}

The authors acknowledge financial support from the Czech Science Foundation (Grant P204/14/30062S), CALIPSO, the DFG under SPP1666, Villum fonden and the Aarhus University Research foundation. J.H. acknowleges the Purkyn\v{e} Fellowship of the ASCR.

\end{acknowledgments}

\bibliography{Bi2Se3library}

%%%%%%%%%%%%%%% END DOCUMENT %%%%%%%%%%%%%%%%%%%%%%%%%%%%%%%%%%%%%%%%%%%%%%

\end{document}